# AlN-buffered superconducting NbN nanowire single-photon detector on GaAs

E. Schmidt, K. Ilin, M. Siegel

*Abstract*—We investigated the suitability of AlN as a buffer layer for NbN superconducting nanowire single-photon detectors (SNSPDs) on GaAs. The NbN films with a thickness of 3.3 nm to 20 nm deposited onto GaAs substrates with AlN buffer layer, demonstrate a higher critical temperature, critical current density and lower residual resistivity in comparison to films deposited onto bare substrates. Unfortunately, the thermal coupling of the NbN film to the substrate weakens.
SNSPDs made of 4.9 nm thick NbN films on buffered substrates (in comparison to detectors made from NbN films on bare GaAs) demonstrate three orders of magnitude lower dark count rates and about ten times higher detection efficiency at 900 nm being measured at 90% of the critical current. The system timing jitter of SNSPDs on buffered substrates is 72 ps which is 36 ps lower than those on bare substrate. However, a weaker thermal coupling of NbN nanowire to the buffered substrate leads to a latching effect at bias currents > 0.97 $I_C$.

*Index Terms*— AlN, NbN, GaAs, Superconducting photodetectors, SNSPD, Superconducting thin films

## I. Introduction

Superconducting nanowire single-photon detectors (SNSPDs) are the detectors of choice for quantum-photonic integrated circuits [1]-[3], due to their high detection efficiency (93% [4]), single photon sensitivity, low dark count rate and very good time resolution with timing jitters below 19.8 ps [5]. A complete quantum-photonic circuit consists of sources of indistinguishable photons, a photonic logic and single-photon detectors which all have to be integrated onto one chip. The material of choice for such integration is GaAs. It has been shown that a self-assembled In(Ga)As quantum dot (QD) on GaAs can emit indistinguishable polarization-entangled photon pairs on demand [6]. The functionality of GaAs ridge waveguides along with NbN-SNSPDs has been demonstrated [3].

The growth of high quality NbN on GaAs is challenging due to a large (about 27%) lattice mismatch between both materials [7]-[9]. The growth of ultrathin NbN films of high quality on traditionally used sapphire and MgO substrates is typically performed at high temperatures up to about 800°C. In case of GaAs substrate, at temperatures exceeding 535°C, surface pits grow due to desorption of oxygen and arsenic, significantly increasing the roughness of the GaAs surface [10] that in turn leads to significant imperfections of deposited films. Furthermore, high temperatures during deposition enhance the inter-diffusion of elements at the film-substrate interface which leads to the formation of layers containing Nb-O and Nb-As compounds with destroyed superconductivity [8]. This reduces the effective thickness of the superconducting NbN and thus leads to a decrease of critical temperature of superconducting transition which is caused by the proximity effect [11]-[13].

Using AlN as buffer layer between GaAs we can overcome several of these challenges. AlN grown in a (002) wurzite lattice has a mismatch of only 8% to NbN [14]. AlN can also grow as a zinc blende lattice, in case of a (100) zinc blende AlN, there is almost no mismatch to NbN [14]. AlN acts as diffusion barrier for oxygen [15] and also presumably prevents desorption of $As_2$. It is transparent in the optical range, rendering it suitable for photonic circuits [16]. Furthermore AlN has a higher thermal conductivity than GaAs (380 $Wm^{-1}K^{-1}$[17] compared to 160 $Wm^{-1}K^{-1}$[18] at 4 K) which could enhance the energy relaxation in thin films and thus increase the speed of detectors. Finally, improvements of the superconducting properties of NbN thin films which were deposited onto $SiO_2$ and soda lime glass substrates due to AlN have been shown [19].

In this paper, we present results of a systematic investigation of the influence of a thin AlN buffer layer on GaAs substrate on the superconducting and normal conducting properties of NbN thin films and on the detection efficiency and timing characteristic of SNSPDs which were made of these films.

## II. Thin Films

### A. Sample Preparation

AlN films were grown on single-side polished (100) GaAs 10 mm × 10 mm substrates. The substrates were placed without the use of thermal glue onto the heater which was kept

This work was supported by DFG project SI704/10-1
E. Schmidt, K. Ilin and M. Siegel are with the Institute of Micro- and Nanoelectronic Systems (IMS), Karlsruhe Institute of Technology, Hertzstrasse 16, Karlsruhe, Germany (e-mail: ekkehart.schmidt@kit.edu)

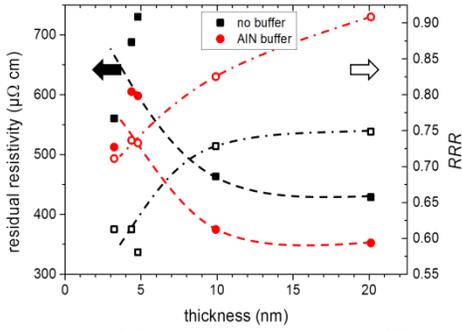

**Fig. 1.** Dependence of the residual resistivity $\rho_n$ (filled symbols) and residual resistance ratio *RRR* (open symbols) on thickness of NbN films which were deposited onto bare GaAs (black squares) and onto AlN-buffered GaAs (red circles) substrates. The dashed lines are a guide to the eye.

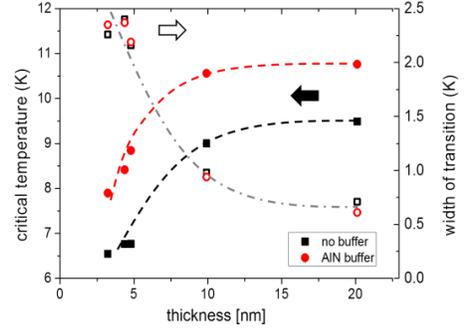

**Fig. 2.** Dependence of the critical temperature $T_C$ (filled symbols) and superconducting transition width $\Delta T_C$ (open symbols) on thickness of NbN films deposited onto bare GaAs (black squares) and on AlN-buffered GaAs (red circles) substrates. The dashed lines are a guide to the eye.

at 500 °C. Prior to deposition, the substrates were pre-cleaned using a low power (20 W) RF-sputtering process in Ar-atmosphere at a pressure of $1\times10^{-2}$ mbar. The AlN deposition was done in-situ, by using reactive DC-magnetron sputtering of an aluminum target in an Ar + $N_2$ atmosphere at a total pressure of $p_{total} = 4.0\times10^{-3}$ mbar with a partial pressure of Ar $p_{Ar} = 3.0\times10^{-3}$ mbar. We deposited 10 nm AlN at a rate of 0.11 nm s$^{-1}$.

NbN films were deposited *ex-situ* without additional pre-cleaning onto bare GaAs and AlN/GaAs substrates simultaneously to provide the best comparability. The NbN films were deposited using reactive DC magnetron sputtering of a pure Nb target in Ar + $N_2$ atmosphere, at a heater temperature of 500 °C. The total pressure was adjusted to $p_{total} = 2.2\times10^{-3}$ mbar with a partial pressure of Ar $p_{Ar} = 1.9\times10^{-3}$ mbar. The deposition rate was 0.10 nm s$^{-1}$. We deposited films with thicknesses, $d$, ranging from 3.2 to 20 nm.

All as-deposited films were patterned into single-bridge structures of different widths between 1 and 10 μm and 20 μm length using photolithography and reactive ion etching technique. Results which will be presented and discussed below are averaged over 10 samples for each thickness of the NbN films.

*B. Film characterization*

We determined superconducting and normal conducting properties of NbN films using a four-terminal sensing measurement scheme. The samples were cooled down in a liquid helium Dewar by submerging them into the liquid helium.

The resistivity $\rho(T)$ of all studied films increases with decreasing temperature, reaching a maximum value $\rho_n$ (we associate this resistivity with the residual resistivity of our films) then the films start transition into superconducting state. The residual resistivity of the measured sample increases with decreasing thickness (Fig. 1). The residual resistance ratio *RRR* was calculated as a ratio of the resistivity at 300 K to $\rho_n$ and increases with decreasing $d$. The $\rho_n$ of the NbN films which were deposited onto AlN-buffered GaAs is ~15% smaller than the residual resistivity of films on bare GaAs. This correlates with the ≈ 20% larger *RRR* of the NbN/AlN/GaAs systems.

The critical temperature $T_C$ (Fig. 2) was determined as the temperature at which the resistance of the sample reaches 0.1% of its residual resistivity $\rho_n$. Independently of the type of substrate, $T_C$ decreases with decreasing $d$, for thicknesses smaller than about 10 nm. The dependence of $T_C$ on thickness of thicker films is much weaker. In Fig. 2 it is seen that the $T_C$ of NbN films on AlN-buffered GaAs substrates is higher than the $T_C$ of the films deposited onto bare substrates in the whole range of studied thicknesses and reaches $T_C = 10.8$ K for the thickest film. We have also found that a spread of $T_C$ values among bridges made from the same film is 0.1 K for bridges made of NbN films on buffered substrate, which is more than two times smaller than the $T_C$ spread of NbN films on bare GaAs. This indicates a better uniformity of NbN films on AlN-buffered GaAs over a large area of 100 mm². The width of the superconducting transition, $\Delta T_C$, was determined as the difference between temperatures which correspond to 10% and 90% of the residual resistivity $\rho_n$ of films on transition from normal to superconducting state. Fig. 2 shows, that with increasing thickness $\Delta T_c$ decreases. In contrast, to the above discussed parameters no influence of type of substrate on the width of the superconducting transition has been observed.

The critical current, $I_C$, and the hysteresis current, $I_H$, were determined from current-voltage characteristics which were measured for all bridges at 4.2 K in current-bias mode. The critical current was determined as the current at which the bridge jumps from superconducting into normal state during an increase of the applied current. The hysteresis current was determined as the current, at which a bridge returns from the resistive to superconducting state by decrease of the applied current. Nominal densities of these characteristic currents ($j_C$ and $j_H$) were calculated using the measured $I_C$ and $I_H$ values and know geometry of the correspondent bridge. The critical current density $j_c(4.2$ K$)$ (see Fig. 3) is ≈ 73% higher for NbN films on AlN/GaAs for all thicknesses. Contrary, the hysteresis current density, which is shown in the same graph in Fig. 3, was found almost unaffected by the presence of the buffer layer. The coefficient, $B$, describing the thermal coupling of the superconducting film to the substrate has been found from a fit of $j_H(T)$ by the following equation [20]:

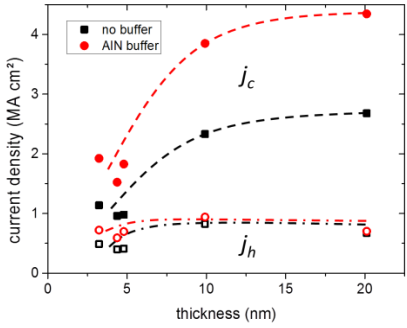

**Fig. 3.** Dependence of density of critical (filled symbols) and hysteresis (open symbols) current on thickness of NbN films on bare GaAs (black squares) and on AlN-buffered GaAs (red circles) substrate. The dashed lines are a guide to the eye.

$$j_h = \left[ \frac{B}{4d\rho_n} \cdot (T_c^4 - T_b^4) \right]^{0.5} \quad (1)$$

where $T_b$ is the bath temperature. The thermal coupling coefficient for GaAs substrates without buffer layer is $B = 2000$ Wm$^{-2}$K$^{-4}$ for thicknesses of NbN films $d > 10$ nm. For smaller thicknesses the $B$ value decreases to 1140 Wm$^{-2}$K$^{-4}$ at $d = 3.3$ nm. On buffered substrates $B$ is almost constant for all measured thicknesses, about 1000 Wm$^{-2}$K$^{-4}$ which is two times smaller for thick films but only 14% smaller for the thinnest films with d ≤ 5 nm.

*C. Discussion and Analysis*

NbN films deposited onto AlN-buffered substrates in comparison to films on bare GaAs are characterized by a) higher critical temperature (Fig. 2), b) lower residual resistivity and higher *RRR* (Fig. 1), c) larger values of the critical current density at 4.2 K, d) almost unchanged width of the superconducting transition (Fig. 2) and e) weaker thermal coupling of the NbN film to the substrate.

The listed above results can be explained if we assume that the AlN buffer layer, which was deposited onto GaAs using DC reactive magnetron sputtering is not ideal (mono-crystal with well-defined orientation) in term of its crystal structure but polycrystalline or rather amorphous. This surface quality is not too different from the quality of bare GaAs which is usually covered by amorphous oxide layer. In this situation an epitaxial growth of NbN film (typically observed in case of NbN films deposited onto heated sapphire substrate [21]) is not possible. Instead we should expect a more polycrystalline-to-amorphous structure of the NbN film which will be grown on top of this AlN-buffer. This assumption is supported by the observed independence of the width of superconducting transition of the presence of this buffer layer (Fig. 2). At the same time the AlN buffer works as diffusion-stop layer between the oxidized surface of the GaAs substrate and the NbN film and thereby protects the superconductor against oxidation of the to the substrate adjacent surface. In frame of the intrinsic proximity effect model [12] this reduction of the oxidized layer, which is with a suppressed superconducting strength, leads to an increase of the critical temperature of thin superconducting films (Fig. 2). Furthermore, oxides of Nb have a higher resistivity than "pure" Nb or NbN. Therefore, a deposited NbN film can be considered as a parallel connection of three layers with different resistivity: a superconducting body of the film with lower resistivity which is sandwiched between two layers with higher resistivity (a layer adjacent to the substrate and the surface layer of film). The total effective resistivity of the film is a correspondent sum of contributions of these layers. A decrease of resistivity of the surface layer in the NbN film, which is adjacent to the AlN buffer, leads to a decrease of resistivity of the whole NbN film and an increase of *RRR* (Fig. 1). Amorphous materials are characterized by a lower thermal conductivity in comparison to single-crystalline samples of the same material. Therefore the weakening of the thermal coupling can be explained by a polycrystalline or amorphous structure of the AlN-film compared to the single-crystalline GaAs. The presence of an amorphous AlN layer with a low thermal conductivity is instead of leading to an enhancement of the cooling efficiency of the NbN film, leading to a worsening of thermal contact between films and substrate as it was shown at the end of section IIB.

### III. SNSPDs

*A. Patterning*

The SNSPDs were patterned from 4.9 nm thick NbN films which were deposited onto bare GaAs and AlN-buffered substrates at a temperature of the heater of 600°C. Due to the increased heater temperature the $T_c$ of the NbN film was increased also. The as-deposited films have a $T_c$ of 9.0 K on bare GaAs and 10.3 K on AlN-GaAs substrate. Nanowires with a width of 120 nm and a length of 218 µm were patterned as spiral to reduce current crowding [22] and enhance critical current $I_c$ and therefore detection efficiency *DE* of the detector [23]. The critical current, which was measured at $T = 4.2$ K, was 8.8 µA for detectors on bare GaAs and 16.5 µA on AlN-buffered GaAs substrates, correspondently.

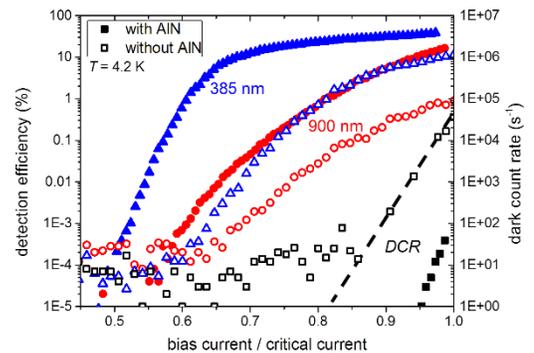

**Fig. 4.** Dependence of the detection efficiency *DE* at a wavelength of $\lambda = 900$ nm (red circles) and $\lambda = 385$ nm (blue triangles) and the dark count rate *DCR* (black squares) on the normalized bias current. Open symbols show the SNSPD without AlN buffer layer, closed symbols the SNSPD with buffer layer. The dashed line is an exponential fit to the *DCR* above 0.85 $I_b$ for the SNSPD directly on GaAs to separate intrinsic dark counts [25] from electronic noise.

*B. Detection efficiency, DCR and timing jitter*

The detectors were characterized using a vacuum dipstick submerged in a liquid helium Dewar. The detector was electrical connected using a low-temperature biasT. A battery-powered low-noise current source was used to bias the detector. The detector pulses were amplified at room temperature using a gain of 57 dB and a bandwidth of 2.9 GHz. Optical radiation was coupled to the detector by a multimode optical fiber.

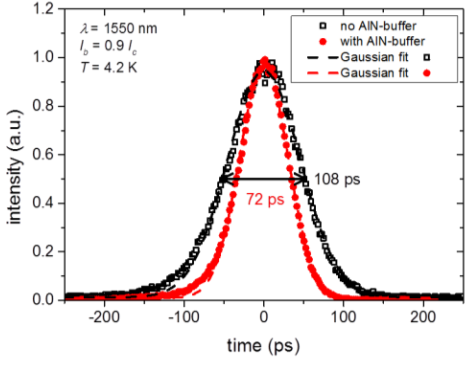

**Fig. 5.** Timing histograms of the system timing jitter. Open black squares show the detector on GaAs, red filled circles the Detector with buffer layer. Given jitter values are the full-width-half-maximum values (FWHM) for the Gaussian fit.

We investigated the bias dependence of the detection efficiency, *DE,* and dark count rate, *DCR,* of our detectors at two different wavelengths. We used a LED light source at a wavelength of 385 nm; photons with $\lambda = 900$ nm were filtered out of the broadband radiation of a thermal light source. Intensities of radiation at both wavelengths were damped down to single-photon level. We used $\lambda = 900$ nm since it is a typical wavelength for photonic circuits on GaAs with embedded InGaAs quantum dots [24],[3]. The detection efficiency at each wavelength was calculated as:

$$DE = \frac{N_{counts} - N_{darkcounts}}{N_{photons}}, \quad (2)$$

where $N_{counts}$ is the number of registered counts, $N_{photons}$ is the number of photons reaching the active area of the SNSPD and $N_{darkcounts}$ the number of measured dark counts.

The dependencies of *DE* and *DCR* of both detectors at both wavelengths on the relative bias current, $I_b/I_C$, are shown in Fig. 4. The detector on bare GaAs substrate has a three orders of magnitude higher *DCR* than the detector on AlN-buffered substrate. Electronic noise can be seen in Fig. 4 at bias currents below $0.85I_C$ for the detector on bare GaAs. This is because output pulses of this detector were small due to lower critical current and correspondingly smaller bias currents. Therefore the set voltage trigger level was just above the noise level of our readout chain leading to contribution of electronic noise into measured *DCR*. Consequently the intrinsic *DCR* for this SNSPD was separated from electronic noise by an exponential fit of the *DCR* above $0.85I_C$. This can be done since the intrinsic *DCR* has an exponential dependence [25] and the dark counts at high bias currents are several orders of magnitude higher than the electronic noise. At $\lambda = 385$ nm, the *DE* at $I_b > 0.75I_C$ for the SNSPD on AlN-buffered substrate starts leveling out, though it is not fully saturated and shows some gradual increase from $DE = 19\%$ at $I_b = 0.75I_c$ to 31% at $I_b = 0.9I_c$ since the detector does not have perfect edge roughness and with further increasing bias current the area of detector with ultimate *DE* grows. At $I_b = 0.9I_c$ the *DE* is six times higher than the *DE* of the detector on bare GaAs. At $\lambda = 900$ nm the *DE* of the SNSPD on AlN-buffered substrate is by an order of magnitude larger than the *DE* of the detector on bare GaAs and about 5% at $0.9I_C$. We explain the increase in *DE* with the better quality and higher uniformity of the NbN film and consequently of the nanowires on GaAs with AlN-buffer layer.

The system timing jitter was measured using an Er-doped fiber based fs-laser with a pulse length < 150 fs and a repetition rate of 100 MHz at a wavelength of 1550 nm [26]. We used the trigger output pulse of the laser as a reference and measured the timing distribution of the SNSPD response using a 32-GHz real-time oscilloscope [27]. The full-width-half-maximum (FWHM) timing jitter was determined out of the Gaussian fit of the measured pulse time distribution Fig. 5. The effective timing jitters for the detectors are lower: our measurement also includes the jitter of the used laser, the RF-readout circuit and the internal jitter of the oscilloscope. The system jitter of the detector on AlN-buffered GaAs substrate is 72 ps which is about 36 ps lower than the system timing jitter of the detector on GaAs without buffer layer. Since the timing jitter is dependent on the absolute value of the bias current of a SNSPD [28], the smaller system timing jitter of the detector on buffered substrate correlates with the higher critical current of this SNSPD in comparison to detectors on bare GaAs substrate.

We have to note that the detector on AlN-buffered GaAs starts latching at $I_b > 0.97I_C$. This latching was not observed for the SNSPD on bare GaAs. A possible reason of latching of the detector made on buffered substrate can be the worsened thermal coupling of the NbN films to the substrate which is caused by the presence of an amorphous AlN buffer layer.

## IV. CONCLUSION

We have shown that AlN buffer layers deposited by reactive magnetron sputtering onto heated GaAs substrates result in the modification of superconducting and normal state properties of NbN films. The films are characterized by about 1.3 K higher $T_C$, 73% higher $j_C(4.2 \text{ K})$, lower resistivity, larger *RRR*, and weaker thermal coupling to GaAs substrate. At the same time the width of superconducting transition of NbN films on buffered substrates has been found almost unchanged in comparison to NbN films which were deposited onto bare GaAs substrates. Therefore, we can conclude that the amorphous AlN layer acts as a diffusion barrier for oxygen and prevents niobium to replace the arsenic at the GaAs surface.

SNSPDs, which were made from NbN films deposited onto AlN-buffered GaAs, demonstrate a reduction of dark counts by three orders of magnitude. The detection efficiency of these detectors has been improved by an order of magnitude at a wavelength of 900 nm. We reached a *DE* of the SNSPD on AlN of 31% at $I_b = 0.9I_c$ and $\lambda = 385$ nm. Due to higher critical currents of these detectors, the timing jitter was decreased by 36 ps and amounts to about 72 ps at $I_b = 0.9 I_C$. At bias currents > $0.97 I_C$ latching of detectors on buffered substrates has been observed. We attribute this latching to a weaker thermal coupling of the NbN nanowire to the substrate which is caused by the amorphous AlN buffer layer.

REFERENCES

[1] J. Sprengers, A. Gaggero, D. Sahin, S. Jahanmirinejad, G. Frucci, F. Mattioli, R. Leoni, J. Beetz, M. Lermer, M. Kamp, S. Höfling, R. Sanjines and A. Fiore, "Waveguide superconducting single-photon detectors for integrated quantum photonic circuits", *Appl. Phys. Lett.*, vol. 99, no. 18, p. 181110, 2011.
[2] W. Pernice, C. Schuck, O. Minaeva, M. Li, G. Goltsman, A. Sergienko and H. Tang, "High-speed and high-efficiency travelling wave single-photon detectors embedded in nanophotonic circuits", *Nature Communications*, vol. 3, p. 1325, 2012.
[3] G. Reithmaier, S. Lichtmannecker, T. Reichert, P. Hasch, K. Müller, M. Bichler, R. Gross and J. Finley, "On-chip time resolved detection of quantum dot emission using integrated superconducting single photon detectors", *Sci. Rep.*, vol. 3, 2013.
[4] F. Marsili, V. Verma, J. Stern, S. Harrington, A. Lita, T. Gerrits, I. Vayshenker, B. Baek, M. Shaw, R. Mirin and S. Nam, "Detecting single infrared photons with 93% system efficiency", *Nature Photonics*, vol. 7, no. 3, pp. 210-214, 2013.
[5] J. Toussaint, S. Dochow, I. Latka, A. Lukic, T. May, H. Meyer, K. Il'in, M. Siegel and J. Popp, "Proof of concept of fiber dispersed Raman spectroscopy using superconducting nanowire single-photon detectors", *Opt. Express*, vol. 23, no. 4, p. 5078, 2015.
[6] M. Müller, S. Bounouar, K. Jöns, M. Glässl and P. Michler, "On-demand generation of indistinguishable polarization-entangled photon pairs", *Nature Photonics*, vol. 8, no. 3, pp. 224-228, 2014.
[7] F. Marsili, A. Gaggero, L. Li, A. Surrente, R. Leoni, F. Lévy and A. Fiore, "High quality superconducting NbN thin films on GaAs", *Superconductor Science and Technology*, vol. 22, no. 9, p. 095013, 2009.
[8] G. Reithmaier, J. Senf, S. Lichtmannecker, T. Reichert, F. Flassig, A. Voss, R. Gross and J. Finley, "Optimisation of NbN thin films on GaAs substrates for in-situ single photon detection in structured photonic devices", *J. Appl. Phys.*, vol. 113, no. 14, p. 143507, 2013.
[9] P. Haas, F. Tran and P. Blaha, "Calculation of the lattice constant of solids with semilocal functionals", *Phys. Rev. B*, vol. 79, no. 8, 2009.
[10] A. Guillén-Cervantes, Z. Rivera-Alvarez, M. López-López, E. López-Luna and I. Hernández-Calderón, "GaAs surface oxide desorption by annealing in ultra high vacuum", *Thin Solid Films*, vol. 373, no. 1-2, pp. 159-163, 2000.
[11] Y. Fominov and M. Feigel'man, "Superconductive properties of thin dirty superconductor–normal-metal bilayers", Phys. Rev. B, vol. 63, no. 9, 2001.
[12] L. Cooper, "Superconductivity in the Neighborhood of Metallic Contacts", *Phys. Rev. Lett.*, vol. 6, no. 12, pp. 689-690, 1961.
[13] R. Schneider, B. Freitag, D. Gerthsen, K. Ilin and M. Siegel, "Structural, microchemical and superconducting properties of ultrathin NbN films on silicon", Crystal Research and Technology, vol. 44, no. 10, pp. 1115-1121, 2009.
[14] C. Stampfl and C. Van de Walle, "Density-functional calculations for III-V nitrides using the local-density approximation and the generalized gradient approximation", *Phys. Rev. B*, vol. 59, no. 8, pp. 5521-5535, 1999.
[15] T. Yeh, J. Wu and W. Lan, "The effect of AlN buffer layer on properties of AlxIn1−xN films on glass substrates", *Thin Solid Films*, vol. 517, no. 11, pp. 3204-3207, 2009.
[16] M. Stegmaier, J. Ebert, J. Meckbach, K. Ilin, M. Siegel and W. Pernice, "Aluminum nitride nanophotonic circuits operating at ultraviolet wavelengths", *Appl. Phys. Lett.*, vol. 104, no. 9, p. 091108, 2014.
[17] G. A. Slack, R. A. Tanzilli, R.O. Pohl and J. W. Vandersande, "The intrinsic thermal conductivity of AlN", *J. Phys. Chem. Solids*, vol. 48, No. 7, pp.641-647, 1987
[18] N. Chaudhuri, R. Wadhwa, P. Tiku and A. Sreedhar, "Thermal Conductivity of Gallium Arsenide at Low Temperatures", *Phys. Rev. B*, vol. 8, no. 10, pp. 4668-4670, 1973.
[19] T. Shiino, S. Shiba, N. Sakai, T. Yamakura, L. Jiang, Y. Uzawa, H. Maezawa and S. Yamamoto, "Improvement of the critical temperature of superconducting NbTiN and NbN thin films using the AlN buffer layer", *Superconductor Science and Technology*, vol. 23, no. 4, p. 045004, 2010.
[20] A. Stockhausen, K. Il'in, M. Siegel, U. Södervall, P. Jedrasik, A. Semenov and H. Hübers, "Adjustment of self-heating in long superconducting thin film NbN microbridges", *Superconductor Science and Technology*, vol. 25, no. 3, p. 035012, 2012.
[21] A. Semenov, B. Günther, U. Böttger, H. Hübers, H. Bartolf, A. Engel, A. Schilling, K. Ilin, M. Siegel, R. Schneider, D. Gerthsen and N. Gippius, "Optical and transport properties of ultrathin NbN films and nanostructures", *Phys. Rev. B*, vol. 80, no. 5, 2009.
[22] D. Henrich, P. Reichensperger, M. Hofherr, J. Meckbach, K. Il'in, M. Siegel, A. Semenov, A. Zotova and D. Vodolazov, "Geometry-induced reduction of the critical current in superconducting nanowires", *Phys. Rev. B*, vol. 86, no. 14, 2012.
[23] D. Henrich, L. Rehm, S. Dorner, M. Hofherr, K. Il'in, A. Semenov and M. Siegel, "Detection Efficiency of a Spiral-Nanowire Superconducting Single-Photon Detector", *IEEE Trans. Appl. Supercond.*, vol. 23, no. 3, pp. 2200405-2200405, 2013.
[24] M. Schwartz, U. Rengstl, T. Herzog, M. Paul, J. Kettler, S. Portalupi, M. Jetter and P. Michler, "Generation, guiding and splitting of triggered single photons from a resonantly excited quantum dot in a photonic circuit", *Opt. Express*, vol. 24, no. 3, p. 3089, 2016.
[25] T. Yamashita, S. Miki, K. Makise, W. Qiu, H. Terai, M. Fujiwara, M. Sasaki and Z. Wang, "Origin of intrinsic dark count in superconducting nanowire single-photon detectors", *Appl. Phys. Lett.*, vol. 99, no. 16, p. 161105, 2011.
[26] Menlo Systems C-Fiber
[27] Agilent Infiniium 90000 X-Series
[28] L. You, X. Yang, Y. He, W. Zhang, D. Liu, W. Zhang, L. Zhang, L. Zhang, X. Liu, S. Chen, Z. Wang and X. Xie, "Jitter analysis of a superconducting nanowire single photon detector", *AIP Advances*, vol. 3, no. 7, p. 072135, 2013.